# "Visibility and Training in Open Source Software Adoption: A Case in Philippine Higher Education"

Ryan A. Ebardo[1, 2 +]

[1] College of Computer Studies and Engineering, Jose Rizal University, Philippines
[2] College of Computer Studies, De La Salle University, Philippines

**Abstract.** Open Source Software (OSS) has been widely used in the educational environment largely due to its reduced cost of ownership. While OSS has evolved over the years, challenges exist in its implementation and wide adoption. Foremost among these detriments is the lack of available skills across industries. Since future users of this technology will settle in an organizational ecosystem where proprietary and OSS technologies coexist, it is vital to understand their learning environment where they initially acquire their technology skills. The University implements courses that champion the use of open source technologies within its curricula. However, other courses are also anchored on technologies that are proprietary. This study is based on the premise that training or the learning experience and visibility or the prevalence of OSS in the environment influences its adoption among students. Empirical evidences explore the relationship of visibility and training in the adoption of OSS from the perspectives of students in a Philippine university. A modified Technology Acceptance Model incorporating additional constructs is validated using Partial Least Squares – Structural Equation Model. Results of the study confirms the applicability of TAM in this study, training has positive influence on both perceived ease of use and perceived usefulness and visibility has a positive influence on perceived ease of use towards initial acceptability of students of OSS. Educational implications of the study are discussed, limitations are acknowledged and research frontiers are recommended.

## 1. Introduction

The proliferation of Open Source Software (OSS) across industries is projected to dramatically increase in the coming years[1]. Despite the domination of copyright software in the technology sector, OSS demonstrates strong potential to reduce licensing costs and software delivery time while offering better flexibility [2]. While drivers in the adoption of OSS encompass the technical, social and ideological perspectives, organizations adopt this technology for financial reasons and this can be attributed mainly to its perceived low or sometimes free acquisition costs[1]. Although arguments and issues have been raised against the Open Source phenomenon, technology giants have slowly embraced the idea of the integration of OSS in its current infrastructure and processes resulting to innovation and collaboration within the software industry stakeholders. While some organizations have totally embraced OSS, its interoperability and flexibility to integrate with proprietary software creates an environment where the presence of OSS and proprietary software exists [3].

In developing countries, adoption of OSS can stimulate economic growth. This is largely due to the ideology of independence that the OSS movement advocates against licensing and purchasing costs. In fact, governments around the world have institutionalized its use through policies, technology blueprints at a national level [4]. The role of educational institutions in the economy as the training ground for the

---

[+] Corresponding author. Tel: + 639178019672; fax: +632837204
  *E-mail address*: ryan.ebardo@jru.edu.

workforce is pivotal towards the acceleration of national development. Critical to the wider acceptance of OSS across social sectors is its integration in the educational curricula and teaching pedagogy[6, 7]. The degree of influence in the academic activities varies and dependent on the environment. From the perspective of educators, OSS have offered alternatives to traditional proprietary software used in instructional activities such as word processing, examinations delivery and classroom instruction. For example, there are open source web-based applications that allow connected collaborations to accomplish learning tasks[7] that are designed primarily for students and educators. The perceived lower acquisition costs and independence from proprietary software have primarily attracted academic institutions towards the OSS direction eliminating the barrier of economic detriment in access to technology [8].

This paper explores the relationship of visibility and training in the adoption of OSS in a university in the Philippines through the widely tested and accepted Technology Acceptance Model (TAM) of Davis [9]. The University is one of the pioneers in adopting OSS in its academic instruction. While proprietary software is still closely embedded in its technology curricula, OSS has been successfully implemented in its instruction across all disciplines. This paper contributes to existing knowledge through the exploration of the influence of social norm of conformity, observing others use OSS [10] and the efficacy of training [11] in the adoption of OSS in the context of Philippine higher education. In the next section, prior literature relevant to the use of OSS and its adoption is discussed in furtherance followed by the introduction of the research model, methodology applied, and the results of the study. In the last section, a conclusion is offered through the discussion of the implications, limitations and frontiers for future research directions.

## 2. Review of Related Studies

Extant literature has been offered by scholars to explore the adoption of OSS. At an organizational analysis, a prior study on software-intensive organizations have explored practices involving the implementation of OSS in existing workflows and revealed that more than acquiring the technology, barriers and challenges exist[12]. Educational institutions have also embraced OSS in its processes and the degree of its utilization varies from classroom instruction and management, research and vital business processes [5], [6]. Governments have also championed the movement of OSS. The lowered acquisition costs, relaxed licensing agreements, and interoperability of OSS have been identified as primary drivers towards its adoption [9, 14].

At an individual level of analysis, studies have also evaluated how individuals perceive, accept and use OSS technologies. Several researches have extensively used and extended the Technology Acceptance Model of Davis [9] to adopt to a specific OSS research environment [11, 12]. In an empirical study, TAM was utilized to incorporate the social, individual and social characteristics of users to study the adoption of OSS among users in a developing country [14]. Recent studies have also evaluated TAM as relevant in the study of student's acceptance of learning activities using OSS software applications [16, 17].

While literature exists on the study of OSS acceptance at organizational and individual levels, this paper contributes to existing scholarship by exploring the effect of visibility and training among students in the adoption of OSS. Universities provide the foundation for skills and training, so this study will be relevant to OSS adoption for workforce development and education. In addition, this paper strengthens available limited researches in OSS adoption in the context of developing economies.

## 3. Research Model and Methodology

This paper studies the influence of visibility and training in the adoption of OSS in an educational environment. Fig. 1: Research Model extends the Technology Acceptance Model by Davis [9] and incorporates the constructs of training [11] and visibility [17] derived from prior studies.

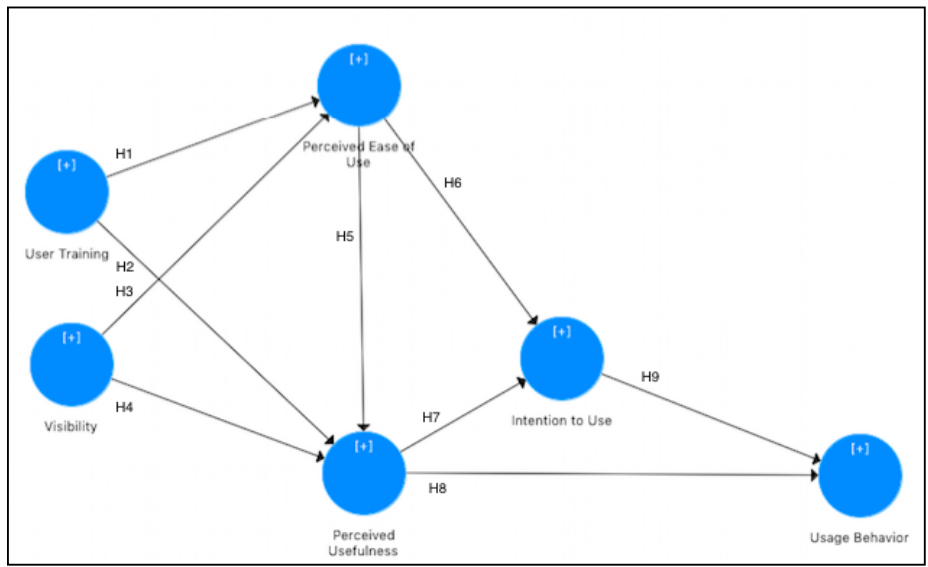
Fig. 1: Research model

As shown in Table 1: Hypotheses Matrix, the following set of hypotheses is drawn which this research will investigate. Studies have revealed that user training positively influence the adoption of OSS. The number of hours devoted and the quality of the training a learner has acquired is directly associated with the perceived ease of use and perceived usefulness of OSS[11]. On the other hand, other scholars have studied the psychological factor of visibility in technology adoption. A study revealed that users demonstrate strong willingness to adopt certain technologies when they see others using them at home, in school or in the workplace [17]. The relationships of perceived ease of use, perceived usefulness, intention to use and usage behavior of the traditional Technology Acceptance Model in the adoption of OSS that were studied by previous researches [12, 17, 19] are tested in the context of Philippine higher education.

Table 1: Hypotheses matrix

| HYPOTHESIS | STATEMENT |
| --- | --- |
| H1 | User Training positively influence Perceived Ease of Use of OSS |
| H2 | User Training positively influence Perceived Usefulness of OSS |
| H3 | Visibility positively influence Perceived Ease of Use of OSS |
| H4 | Visibility positively influence Perceived Usefulness of OSS |
| H5 | Perceived Ease of Use positively influence Perceived Usefulness of OSS |
| H6 | Perceived Ease of Use positively influence Intention to Use OSS |
| H7 | Perceived Usefulness positively influence Intention to Use OSS |
| H8 | Perceived Usefulness positively influenceUsage Behavior towards OSS |
| H9 | Intention to Use positively influence Usage Behavior towards OSS |

To test the research model, a survey instrument was deployed to the students of the University. The questions relative to perceived ease of use, perceived usefulness, intention to use and usage behavior in the instrument were adopted from a similar study [10] while questions to measure the effect of user training and visibility were based from two separate studies [11], [17]. The questions were updated to adapt to the research environment based on the target respondents and the technologies used in the University. Of the 436 respondents, 189 (43%) are females while 247 (57%) are males. In terms of their year levels, 28 (6%) are on their first year, 31 (7%) are on their second year, 172 (40%) are on their third year and 197 (45%) are on their final year as university students. Of the respondents, 8 (2%) students reported that they are on an irregular status. With reference to their disciplines, 175 (40%) are taking up programs in computer studies, 124 (28%) are from business and accounting, 56 (13%) are from hospitality, 49 (11%) are from engineering and 20 (5%) are from education. A total of 12 (3%) students are from other courses.

To exhibit the adequacy of the research instrument, a test to measure convergent validity and measurement reliability was conducted using Partial Least Squares – Structural Equation Model or PLS - SEM through smartPLS. This method is widely used in Information Systems research due to its predictive nature and efficacy in small sample sizes [19]. The results are shown in Table 2: Test of Convergent Validity and Measurement reliability.

Table 2: Test of convergent validity and measurement reliability

| CONSTRUCT | CRONBACH'S ALPHA | COMPOSITE RELIABILITY | AVERAGE VARIANCE EXTRACTED |
|---|---|---|---|
| Intention to Use | 0.968 | 0.976 | 0.891 |
| Perceived Ease of Use | 0.930 | 0.956 | 0.780 |
| Perceived Usefulness | 0.875 | 0.915 | 0.732 |
| Usage Behavior | 0.892 | 0.943 | 0.902 |
| User Training | 0.808 | 0.863 | 0.620 |
| Visibility | 0.901 | 0.937 | 0.833 |

All constructs demonstrated strong evidence of measurement reliability based on Cronbach's Alpha and Composite Reliability tests. The results on both tests are above the threshold of 0.700 ($\alpha>0.700$) demonstrating reliability of all composites. The Average Variance Extracted or AVE test also yielded strong evidence of discriminant validity as evidenced by AVEs that are above 0.500. The instrument is valid and is consistent with other tests using PLS-SEM [17, 19]. The survey was deployed online and was open for thirty (30) days. A total of four hundred thirty six (436) out of four hundred forty seven (447) responses were considered after eliminating invalid entries. All questions used a five-point Likert scale.

## 4. Discussion of Results

To test the modified Technology Acceptance Model, a 1-tailed, bootstrapping technique was performed in the collected data using smartPLS at a significance level of 0.05. Based on the results, this study infers the following: First, perceived ease of use of OSS has a direct, positive relationship with its perceived usefulness (H5) and intention to use (H6). This means that the more the student believes OSS is easy to use, the more the student will think it will be useful and will lead to increased intention to use OSS. Second, perceived usefulness has a direct, positive relationship with intention to use (H7) and usage behavior (H8) towards OSS. The perception of students on how useful OSS is in learning activities leads to intention to adopt and actual usage of OSS technologies. Third, the intention to use positively influence actual usage of OSS (H9). In addition, the students' actual usage of OSS is strongly influenced by their intention to use this technology. All hypotheses (H5, H6, H7, H8, H9) in the Technology Acceptance Model are accepted at a significant level of 0.05 as they have T-Statistics that are higher than minimum threshold [20]. These findings are consistent with previous studies that used TAM in the adoption of OSS [12, 17] and confirms its adaptability to the current research scope. Fourth, user training increases perceived ease of use and perceived usefulness of OSS. The qualities of training that students receive positively influence their perceptions on the ease of use (H1) and usefulness (H2) of OSS in their academic lives. The hypotheses in this construct is accepted at a significant level of 0.05 as the values of T-Statistics (11.336 for H1, 4.4 for H2) are acceptable [20]. The results demonstrate that adequate training is critical to the adoption of OSS to its users [11]. Lastly, visibility leads to perceived ease of use (H3) but has an insignificant effect on the perceived usefulness of OSS from the perspective of the students. On the other hand, the T-Statistics value of H4 (0.015) is rejected, as it did not meet the acceptable threshold. The psychological factor of conformity plays a vital role to the adoption of OSS[17]. As students encounter OSS frequently, their perceptions on its ease of use increase. However, visibility has no direct effect on the perceived usefulness of OSS. This can be attributed to the fact that students enrolled in the University have been oriented on its benefits and usefulness in the various courses where OSS is the core technology in classroom instruction and therefore visibility will have no impact in this relationship. The results are summarized in Table 3: Model Validation.

Table 3: Model validation

| HYPOTHESIS | PATH | PATH COEFFICIENTS | T-STATISTICS | SUPPORTED | SIGNIFICANCE LEVEL |
|---|---|---|---|---|---|
| H1 | User Training → Perceived Ease of Use | 0.675 | 11.336 | Yes | 0.05 |
| H2 | User Training → Perceived Usefulness | 0.327 | 4.400 | Yes | 0.05 |
| H3 | Visibility → Perceived Ease of Use | 0.247 | 3.940 | Yes | 0.05 |
| H4 | Visibility → Perceived Usefulness | 0.001 | 0.015 | No | NA |
| H5 | Perceived Ease of Use → Perceived Usefulness | 0.576 | 7.412 | Yes | 0.05 |
| H6 | Perceived Ease of Use → Intention to Use | 0.662 | 10.389 | Yes | 0.05 |
| H7 | Perceived Usefulness → Intention to Use | 0.260 | 3.855 | Yes | 0.05 |
| H8 | Perceived Usefulness → Usage Behavior | 0.339 | 5.693 | Yes | 0.05 |
| H9 | Intention to Use → Usage Behavior | 0.591 | 9.800 | Yes | 0.05 |

## 5. Conclusion

This study is with limitations. Foremost among them is that the population is confined within the boundaries of a single institution and generalizability is limited. In addition, the questions in the instrument referred only to the existing OSS technologies within the University. This may not be a transparent representation of OSS technologies. The limitations of the study provide opportunities for further research for the scholars. Future research within this domain may use a larger, randomized sample for a complete representation of OSS adoption among students in Philippine higher education and a qualitative approach may be used to verify the findings of this study.

This study contributes to the existing body of knowledge by confirming the validity of the Technology Acceptance Model in the study of the adoption of Open Source Software from the perspective of students in the educational environment. The empirical evidences from this study confirms previous studies that user training has a positive influence in the perceived ease of use and perceived usefulness in the adoption of OSS [11]. In addition, visibility has a positive influence on perceived ease of use and is consistent with published findings [17]. The value of training and visibility among students towards the adoption of OSS was validated by this study. Related researches have argued on the difficulty of implementing OSS in organizations. Therefore, it is paramount that educational institutions such as the University should focus on improving instruction quality and institutionalize policies that promote the use of OSS.